\documentclass[structabstract]{aa}

\usepackage{graphicx}
\usepackage{txfonts}
\begin{document}
   \title{High energy emission from the nebula around the Black Widow binary system containing millisecond pulsar B1957+20}


   \author{W. Bednarek$^1$  \& J. Sitarek$^2$
          }

   \institute{$^1$Department of Astrophysics, University of \L \'od\'z,
              ul. Pomorska 149/153, 90-236 \L \'od\'z, Poland\\
              $^2$IFAE, Edifici Cn., Campus UAB, E-08193 Bellaterra, Spain\\
              \email{bednar@astro.phys.uni.lodz.pl; jsitarek@ifae.es}
             }

   \date{Received ; accepted }


\abstract
{The features of pulsed $\gamma$-ray emission from classical and millisecond pulsars indicate that the high energy radiation processes in their inner magnetospheres occur in a similar way. 
In the last decade several TeV $\gamma$-ray nebulae have been discovered around classical pulsars. 
The above facts suggest that $\gamma$-rays should be produced also in the surroundings of millisecond pulsars.}
{We discuss a model for the bow shock nebula around the well known Black Widow binary system containing the millisecond pulsar B1957+20. This model predicts the existence of a synchrotron X-ray and inverse Compton $\gamma$-ray nebula around this system. We want to find out whether $\gamma$-ray emission from the nebula around B1957+20 could be detected by the future and present Cherenkov telescopes.}
{Using the Monte Carlo method we followed the propagation of relativistic electrons in the vicinity of the pulsar. 
We calculated the very high energy radiation produced by them in the synchrotron process and the inverse Compton scattering of the Microwave Background Radiation and of the infrared radiation from the galactic disk. We also computed the  X-ray emission produced by the electrons in the synchrotron process.}
{We show that the hard X-ray tail emission observed from the vicinity of B1957+20 can be explained by our model. Moreover, we predict that the TeV $\gamma$-ray emission produced by the electrons in the inverse Compton process should be detectable by the future Cherenkov Telescope Array and possibly by the long term observations with the present Cherenkov arrays such as MAGIC and VERITAS. The $\gamma$-ray emission from B1957+20 is expected to be extended,
inhomogeneous, and shifted from the present location of the binary system by a distance comparable to the radius of the nebula.}
{}

\maketitle
%

%
%
\section{Introduction}

PSR B1957+20 was the first millisecond pulsar (MSP) discovered within the binary system belonging to the class of Black Widows (Fruchter et al.~1988). This pulsar has a very small mass companion ($\sim 0.022M_\odot$, van Paradijs et al.~1988) which evaporates under the irradiation from the pulsar magnetosphere. The pulsar has the period of 1.607 ms, the surface magnetic field of $\sim 10^8$ G, and the rotational energy loss rate of $7.5\times 10^{34}$ erg s$^{-1}$.
The distance to the binary system is estimated on 2.5 kpc (from the model for Galactic electron density) consistent with the recently established lower limit $\sim$2 kpc (van Kerkwijk et al.~2011). The binary system is compact with the orbital radius of $1.5\times 10^{11}$ cm. 
The companion star has the radius $\sim 10^{10}$ cm and the surface temperature which varies between 2900 K for the unilluminated side to 8300 K for the illuminated side (Fruchter et al.~1995, Reynolds et al.~2007). Therefore, stellar radiation is not expected to create a very strong target for relativistic particles within the binary system. At present, the companion star loses mass at a rather low rate estimated on $10^{-10}$ M$_\odot$ yr$^{-1}$ (Takata et al.~(2012).

The importance of the high energy processes in the vicinity of PSR B1957+20
has become clear with the discovery of an H$\alpha$ emission nebula (Kulkarni \& Hester~1988).
This emission is expected to be produced in shocks formed in the interaction of the pulsar wind with the interstellar medium. A clear bow shock has been detected which  apex is located at the distance of $\sim0.02\,$pc from the pulsar. The bow shock appears due to the motion of the binary system with the velocity 220 km s$^{-1}$ through the interstellar medium (Arzoumanian et al.~1994).
The X-ray emission has been also reported from the direction of the binary system in the observations of {\it Chandra} (Stappers et al.~2003, Huang \& Becker~2007, Huang et al.~2012). This emission comes from the interior of the bow shock creating a tail behind the moving binary system. The length of the tail is $\sim 10^{18}$ cm (Huang et al.~2012).
The X-ray emission is well described by a single power law spectrum with the index in the range $2.3-2.6$ depending on the absorption model. The extended X-ray feature has been interpreted as emission from energetic electrons which radiate on the crossing time scale of this region by the pulsar moving with velocity of $220$ km s$^{-1}$ (Cheng et al.~2006). 

The Black Widow binary system containing B1957+20 was claimed in the past to be a GeV-TeV $\gamma$-ray source (Brink et al.~1990). But this early report was not confirmed in the analysis of the EGRET data (Buccheri et al.~1996). 
In fact, such high energy emission has been suspected already since the discovery of Black Widow pulsars as a result of either the acceleration of particles within the binary system or in the the shock waves of the pulsar wind (e.g. Arons \& Tavani~1993, Cheng et al.~2006, Takata et al.~2012).
Recently, a pulsed GeV emission from the pulsar B1957+20 has been discovered by {\it Fermi} (Guillemot et al.~2012). The pulsed spectrum is flat above 0.1 GeV (spectral index close to 2) and extends up to $\sim 4$ GeV. The phasogram (light curve folded with the period of the pulsar) shows two well separated peaks. Such structure is also common in the case of classical pulsars. Therefore, it is expected that the radiation processes in the inner magnetosphere of the millisecond pulsar B1957+20 are similar to those occurring in the case of classical pulsars.  This strongly indicate that also processes of acceleration of particles in the pulsar wind are expected to occur similarly. 
Very recently, Wu et al.~(2012) reports detection of orbital modulation of the $\gamma$-ray emission at energies above $\sim$2.7 GeV from the Black Widow pulsar PSR B1957+20. This emission is expected to be produced by electrons from the pulsar wind which comptonize stellar radiation.

We investigate the radiation processes in the supposed pulsar wind nebula around the binary system containing PSR B1957+20. The synchrotron X-ray and inverse Compton (IC) $\gamma$-ray emission is calculated from such nebula for the range of likely parameters. Based on the comparison of the calculated synchrotron spectrum with the observed X-ray emission we conclude on the detectability of the TeV $\gamma$-ray emission from the bow shock nebula surrounding PSR B1957+20.

\section{The nebula around binary system downstream of the bow shock}

\begin{figure}
\vskip 8.5truecm
\includegraphics{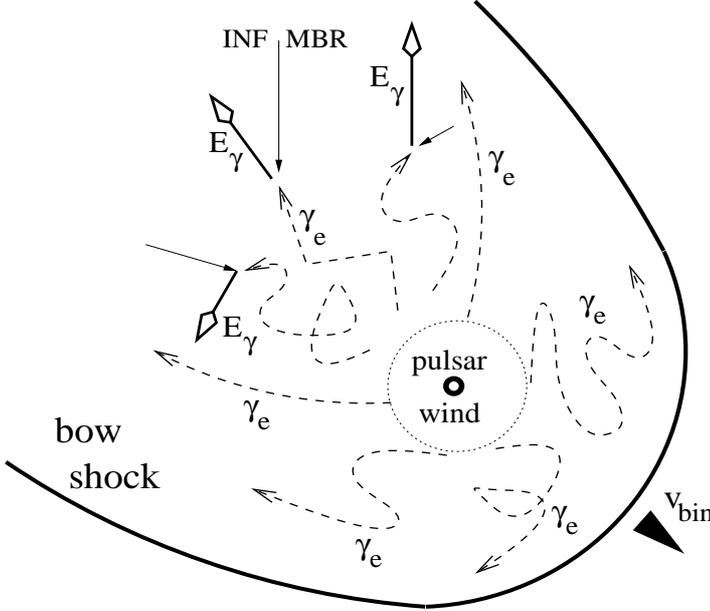}
\caption{Schematic representation of the bow shock nebula around binary system containing the millisecond pulsar B1957+20. The bow shock is created due to the motion of the binary system through the interstellar space with the velocity of $\sim 220$ km$\,$s$^{-1}$. Relativistic electrons with the Lorentz factors $\gamma_{\rm e}$ are accelerated by the pulsar itself or by the shocks due to the pulsar wind interactions. 
The electrons are collimated by the bow shock in the direction opposite to the motion of the binary system. These electrons comptonize the Microwave Background Radiation (MBR) and the infrared radiation (INF) from the galactic disk. As a result $\gamma$-ray photons are produced (tagged as $E_\gamma$) at the region behind the pulsar.}
\label{fig1}
\end{figure}
Since the proprieties of  high energy $\gamma$-ray emission from the millisecond pulsars and classical radio pulsars are surprisingly similar (see the first pulsar catalogue, Abdo et al.~2010), it seems clear that the processes occurring in their inner magnetospheres are these same. Therefore, millisecond pulsars should also produce relativistic pulsar winds with the parameters similar to those observed around classical pulsars.
However, nebulae around MSPs are  expected to have a very complicated structure (and also other proprieties) since many MSPs form compact binary systems which additionally move in the interstellar space with large velocities. In fact, this is the case of the binary system PSR B1957+20. The pulsar wind around B1957+20 is expected to interact with the 
induced wind of the low mass companion star within a small solid angle, of the order of $\sim$0.01 sr, corresponding to eclipse time 
of the pulsar radio emission by the wind of the companion star 
(Fruchter et al.~1988). Therefore, most of the pulsar wind is expected to escape unaffected from the binary system. Due to the fast velocity of the binary system, the pulsar wind has to interact with the interstellar medium creating a bow shock. Such bow shock has been detected in H$\alpha$ emission in the case of B1957+20. The distance of the apex of the bow shock to the pulsar is estimated on $\sim 0.02$ pc (Kulkarni \& Hester~1988). This bow shock confines the pulsar wind 
at least in the direction of pulsar's motion. Relativistic electrons in the wind can diffuse mainly in the direction opposite to the pulsar's motion. Cheng et al.~(2006) have proposed that the synchrotron radiation from such ultrarelativistic electrons is responsible for the observed X-ray tail extending along the axis of the bow shock in the direction opposite to the pulsar velocity. We intend to perform calculations of the Inverse Compton (IC) $\gamma$-ray emission from such relativistic electrons applying  general scenario proposed by Cheng et al.~(2006).
These authors argue that efficient synchrotron emission by electrons can occur on the dynamical time scale of the pulsar crossing the length of the tail estimated on $\sim$10$^{18}$ cm (Huang et al. 2012). This dynamical time scale is equal to, 
\begin{eqnarray}
\tau_{\rm dyn} = R/v_{\rm bin}\approx 1.5\times 10^{11}R_1~~~{\rm s},
\label{eq1}
\end{eqnarray}
\noindent
where the velocity of the binary system is $v_{\rm bin} = 220$ km s$^{-1}$ and the length of the tail is $R = 1R_1$\,pc. Applying the observed length of the synchrotron emission, we can estimate the  optical depth for electrons on the IC scattering of the Microwave Background Radiation (MBR) in the Thomson regime 
(true for electrons with energies below $\sim 100$ TeV) on,
$\tau = c\tau_{\rm dyn}n_{\rm MBR}\sigma_{\rm T}\approx 0.35$, where $c$ is the velocity of light, $\sigma_{\rm T}$ is the Thomson cross section, and $n_{\rm MBR}$ is the photon density of the Microwave Background Radiation.
Note however that electrons cool only partially in the region of observed X-ray emission. Many of them escape from this region but continue to interact with the MBR and other soft photon field, producing high energy $\gamma$-rays.
Therefore, we expect the appearance of the $\gamma$-ray nebula in the vicinity of the Black Widow binary pulsar. This nebula should be shifted in respect to the observed location of the binary system in the direction opposite to the pulsar's motion. 

On the other hand, the energy loss time scale of electrons on the IC scattering in the Thomson regime is,
\begin{eqnarray}
\tau_{\rm IC}^{\rm T} = m_{\rm e}c^2\gamma_{\rm e}/(4cU_{\rm rad}\sigma_{\rm T}\gamma_{\rm e}^2/3)~~~{\rm s},
\label{eq2}
\end{eqnarray}
\noindent
where $U_{\rm rad}$ is the energy density of the soft radiation field
equal to $0.3$ eV cm$^{-3}$ for the Microwave Background Radiation (MBR) and to $\sim 1.5$ eV cm$^{-3}$ for the infrared radiation with characteristic energies $\sim 0.01$ eV, produced in the galactic disk (e.g. see the values calculated in Hui et al.~(2011) based on the
GALPROP code developed by Strong \& Moskalenko~1998), and $m_{\rm e}$ is the rest mass of  an electron. For these energy densities we obtain the energy loss time scales of the order of $\tau_{\rm IC}^{\rm T}\sim 6.3\times 10^{19}/\gamma_{\rm e}$ s for the MBR and $\sim 1.3\times 10^{19}/\gamma_{\rm e}$ s for the infrared radiation, where $\gamma_{\rm e}$ is the Lorentz factor of the electrons. 
In order to cool the electrons efficiently on the IC process during the dynamical time of the moving pulsar, the emission region should have the diameter of the order of $R\approx 8.5\times 10^7/\gamma_{\rm e}$ pc. For example, in the region of 10 pc, electrons with energies larger 
than  $\sim 4$ TeV (but below $\sim 100$ TeV since the electrons have to interact in the Thomson regime) should be able to produce efficiently $\gamma$-rays in the IC process by scattering infrared photons from the galactic disk.
Note that, the region of the $\gamma$-ray production in the IC process should be clearly shifted from the pulsar position in the direction of the observed tail X-ray emission. This region should be also inhomogeneous with 
higher energy $\gamma$-rays produced closer to the pulsar. 

In the above estimates we neglected the energy density of stellar photons, in respect to the MBR and infrared radiation 
at the region of the acceleration of electrons (the shock in the pulsar wind).
In fact, the energy density of stellar photons depends on the distance from the star as $U_\star\approx 4.5\times 10^{-5}/D_{\rm 18}^2$ eV cm$^{-3}$, where the distance from the star is $D = 10^{18}D_{18}$ cm. It is assumed that the companion star in the binary system PSR 1957+20 has the radius $10^{10}$ cm and most of its surface has temperature close to $\sim$3000 K (Fruchter et al.~1995). For these parameters, the electron energy losses are dominated by scattering of the infrared photons for distances above $\sim 5\times 10^{15}$ cm.
Note also that the scattering of the optical photons from the star occurs in the Klein-Nishina regime for electrons with energies
above $\sim 100$ GeV. Therefore, the effects of scattering stellar radiation by the TeV electrons can be safely neglected.

The region of the $\gamma$-ray production can be also affected by the diffusion of the electrons in the pulsar wind downstream of the pulsar wind shock. We estimate the diffusion distance of the electrons, as a function of their energy, and compare it with the time scale corresponding to the dynamical motion of the pulsar. For the Bohm diffusion approximation,  the diffusion distance is $R_{\rm dif} = \sqrt{2D_{\rm dif}t}$, where $D_{\rm dif} = cR_{\rm L}/3$ is the diffusion coefficient,
$R_{\rm L}$ is the Larmor radius of electrons, $B$ is the magnetic field strength in the considered region, and $t$ is the diffusion time. If $B$ is fixed on $1\mu$G, then $D_{\rm dif}\approx 1.5\times 10^{19}\gamma_{\rm e}$ cm$^2$s$^{-1}$ and $R_{\rm dif}\approx 5.5\times 10^9\sqrt{\gamma_{\rm e}t}$ cm. The spread of the emission region due to the diffusion process is smaller than that one due to the motion of the pulsar, i.e. $R_{\rm dif} < R_{\rm dyn} = v_{\rm pul}t$, for the following condition $t>6.2\times 10^4\gamma_{\rm e}$ s. We compare this condition with the energy loss time scale on the IC process in the Thomson regime (see Eq.~2 and estimates below). It is found that electrons with energies below $\sim$7 TeV lose energy on production of $\gamma$-rays when the ballistic motion of the binary system determines the morphology of the $\gamma$-ray source.
We conclude that depending on the electron energy, the dimension of the $\gamma$-ray 
source is determined either by the motion of the Black Widow binary system through the interstellar medium or by the diffusion process of the electrons.

\section{Relativistic electrons in nebula} 

We estimate the magnetic field strength around the pulsar, above its light cylinder radius, by extrapolating it from the pulsar surface. The magnetic field strength is then given by,
\begin{eqnarray}
B(R)\approx 4.4\times 10^{-6}\sigma^{1/2}B_8/(P_{\rm ms}^2R_{18})~~~{\rm G},
\label{eq3}
\end{eqnarray}
\noindent
where $R = 10^{18}R_{18}$ cm is the distance from the pulsar, $B_{\rm NS} = 10^8B_8$ G is the magnetic field strength on the neutron star surface, $P =10^{-3}P_{\rm ms}$ s is the period of the millisecond pulsar,
and $\sigma$ is the magnetization parameter of the pulsar wind. $\sigma$ has been estimated in the case of the Crab Nebula on 0.003 (de Jager \& Harding~1992) and in the case of the Vela Nebula on $\sim$0.1 (Sefako \& de Jager~2003). $\sigma$ is expected to be in the range $0.001-0.01$ in the modeling of the Crab Nebula presented by Kennel \& Coroniti (1984).
The magnetic field given by Eq.~3, is expected to be enhanced at the shock region in the pulsar wind by a factor of $\sim 3$.  
Downstream of the shock, electrons are isotropized and start to radiate efficiently synchrotron radiation.
Therefore, the magnetic field in the region downstream of the shock is an important factor which determines the diffusion of the relativistic electrons and production of the synchrotron radiation.
The maximum energies to which the electrons can be accelerated in the pulsar shock region can be estimated from,
\begin{eqnarray}
E_{\rm max} = cR_{\rm sh}B(R_{\rm sh})\approx 4\times 10^{6}\sigma^{1/2}B_8/P_{\rm ms}^2~~~{\rm GeV}.
\label{eq4}
\end{eqnarray}
\noindent
Note that this simple formula gives the energies of electrons present in the Crab Nebula comparable to those expected from the modelling of its 
multi-TeV $\gamma$-ray spectrum (e.g. de Jager \& Harding~1992).

As noted above, {\it Chandra} has detected the tail behind the pulsar B1957+20 in the energy range 0.3-8 keV (Huang et al.~2012). If this emission is due to the synchrotron process from the relativistic electrons, then the Lorentz factors of the electrons can be estimated from,
\begin{eqnarray}
\varepsilon = m_{\rm e}c^2(B/B_{\rm cr})\gamma_{\rm e}^2,
\label{eq5}
\end{eqnarray}
\noindent
where $\varepsilon = 8$ keV is the energy of synchrotron photons, $B$ and $B_{\rm cr} = 4.4\times 10^{13}$ G are the magnetic field  in the emission region and the critical magnetic field strength.
The inspection of the above equations allows us to conclude that the production of the synchrotron photons with $\sim$10 keV energies is possible provided that the Lorentz factors of electrons are at least $\gamma_{\rm e} = 
2.2\times 10^8P_{\rm ms}R_{18}^{1/2}/(\sigma^{1/4}B_8^{1/2})$, obtained by substitution of Eq.~3 into Eq.~(5).
Electrons are accelerated to such energies provided that the magnetic field is strong enough, i.e. the shock in the pulsar wind appears close to the pulsar.  For the parameters of PSR B1957+20,
the distance of the shock has to be below $R_{18}\approx 3.4\times 10^3\sigma^{3/2}$, which equals to $R_{\rm sh}\approx 10^{17}-10^{20}$ cm for $\sigma$ in the range $0.001-0.1$. 
This condition is consistent with the observations of the PWNe around classical pulsars. For example, in the case of the Crab Nebula the shock is located at the distance of $\sim 3\times 10^{17}$ cm (Kennel \& Coroniti~1984).

It is not clear at present in what process electrons reach such large energies. This might be reconnection of the magnetic field or the shock acceleration mechanism. In the second case, the limit on the maximum energies of the electrons have to be consistent with the limit due to the presence of the synchrotron energy losses already during the acceleration process.
The maximum energies of the electrons, due to the saturation by the synchrotron energy losses, can be derived from the comparison of the electron acceleration time scale,
\begin{eqnarray}
\tau_{\rm acc}\approx 1E_{\rm e}/(\chi_{-1}B)~~~{\rm s},
\label{eq6}
\end{eqnarray}
\noindent
with the synchrotron energy loss time scale,
\begin{eqnarray}
\tau_{\rm syn} = E_{\rm e}/{\dot E}_{\rm syn}\approx 370/(B^2E)~~~{\rm s},
\label{eq7}
\end{eqnarray}
\noindent
where ${\dot E}_{\rm syn} = (4/3)cU_{\rm B}\sigma_{\rm T}E_{\rm e}^2/m_{\rm e}^2\approx 0.0027B^2E^2$ TeV/s, the acceleration efficiency is parametrised by the factor $\chi = 10^{-1}\chi_{-1}$, and $E_{\rm e}$ is the electron energy in TeV.
Energies of the electrons can not be larger than,
\begin{eqnarray}
E^{\rm max}_{\rm syn}\approx 2\times 10^4({{\chi_{-1}}\over{B}})^{1/2}~~~{\rm GeV}\approx  5\times 10^6{{P_{\rm ms}R_{18}^{1/2}}\over{\sigma^{1/4}B_8^{1/2}}}~~~{\rm GeV},
\label{eq8}
\end{eqnarray}
\noindent
For the pulsar with the parameters of PSR B1957+20, $E_{\rm max}$ is lower than $E_{\rm max}^{\rm syn}$ for the location of the shock at $R_{18} > 0.05 \sigma^{3/2}$, which corresponds to $R_{\rm sh} > 1.6\times 10^{15}$ cm for $\sigma = 0.1$. Therefore, we conclude that for the expected localizations of the shock in the nebula around the pulsar B1957+20 (above $\sim 10^{15}$ cm), the synchrotron energy losses can not limit the acceleration process of the electrons below the maximum possible energies given by Eq.~\ref{eq4}.

\begin{figure*}
\vskip 12.7truecm
\includegraphics{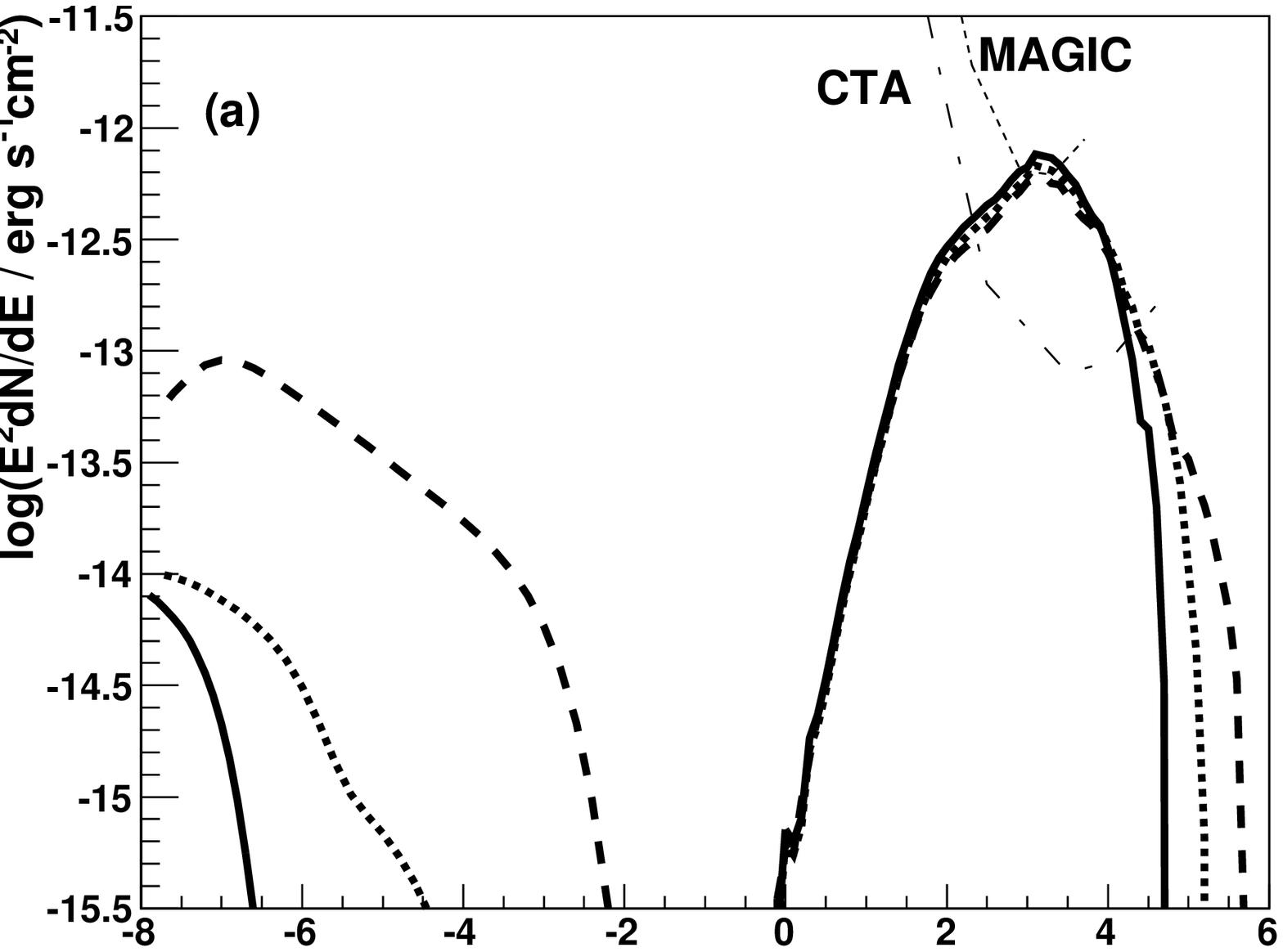}
\includegraphics{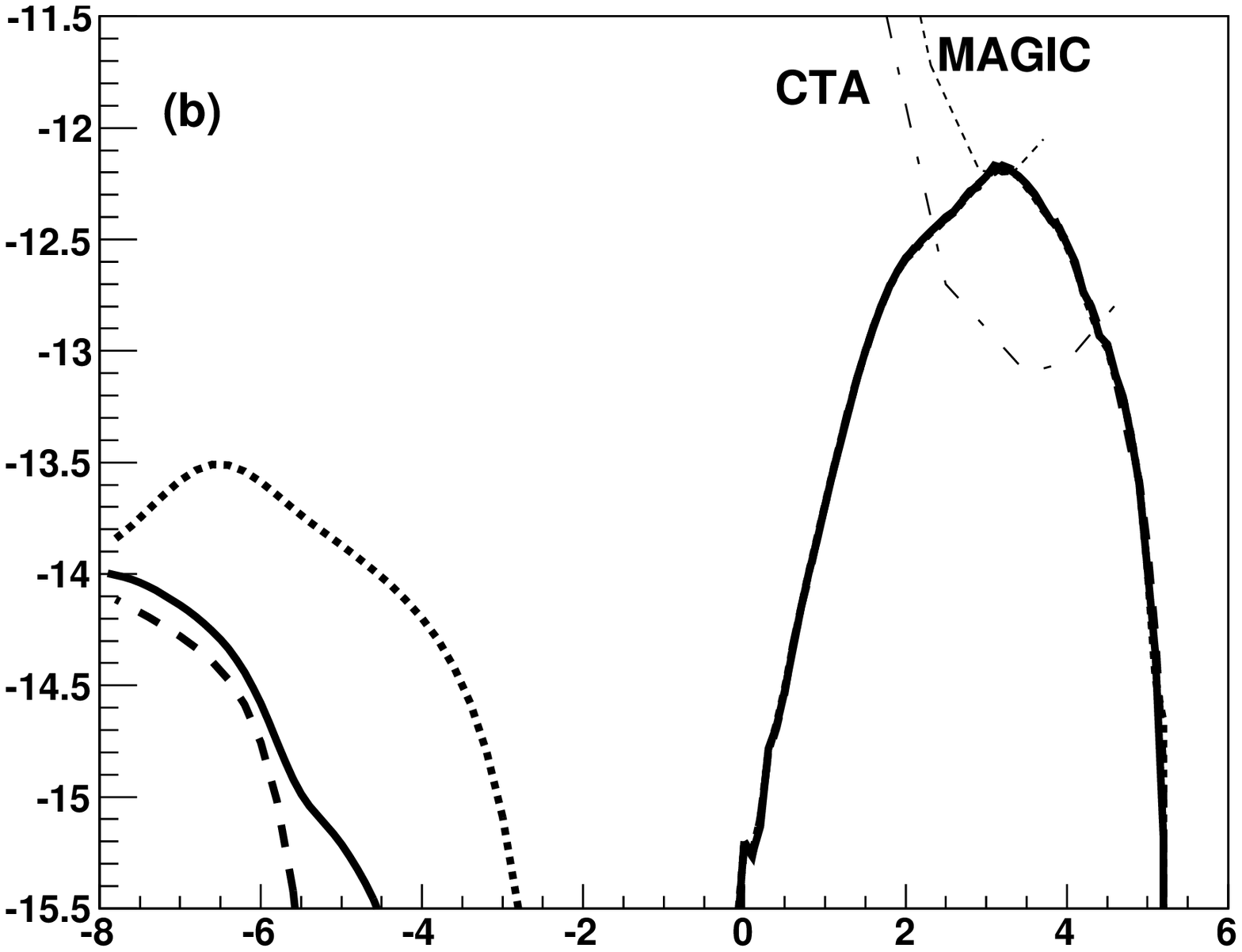}
\includegraphics{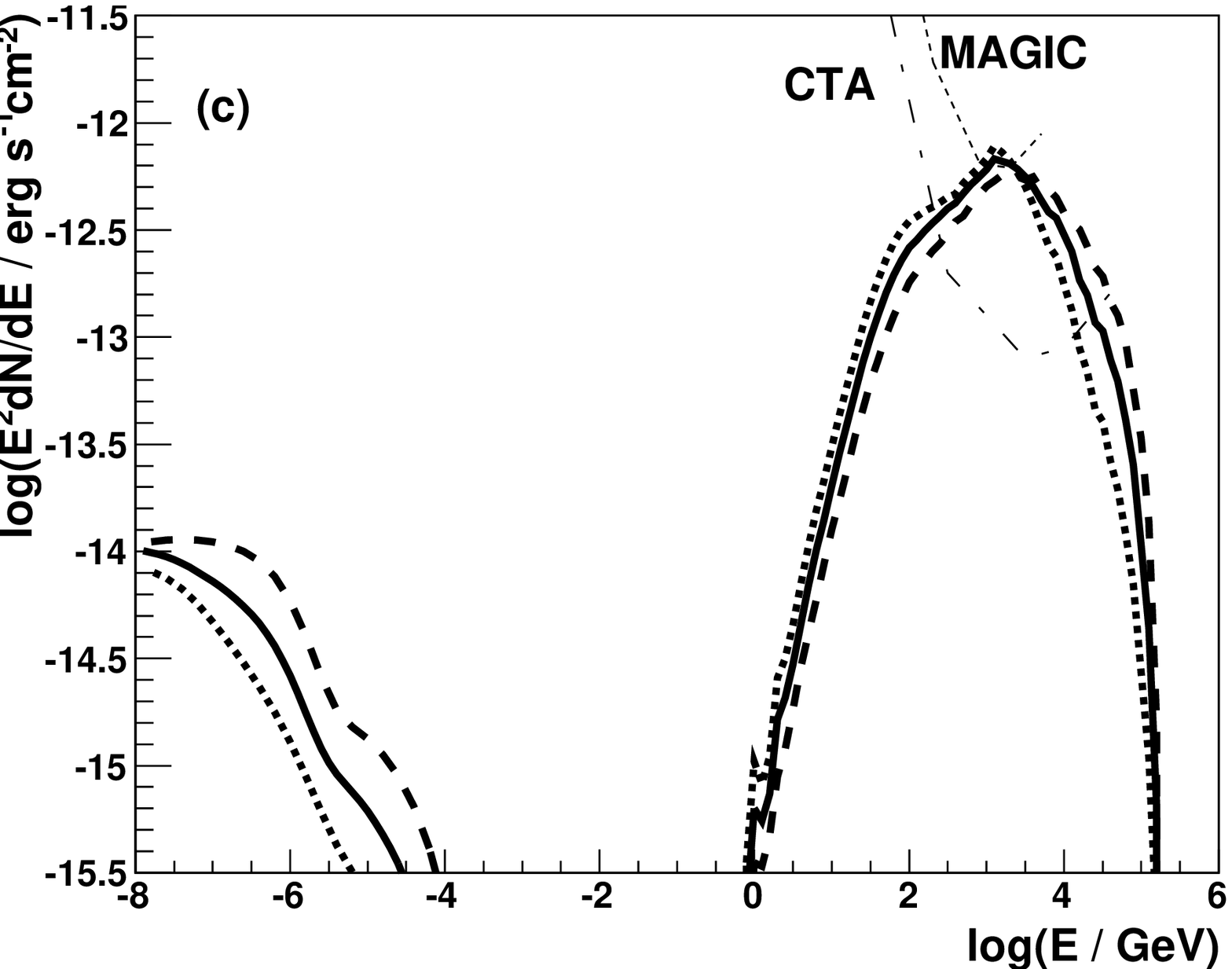}
\includegraphics{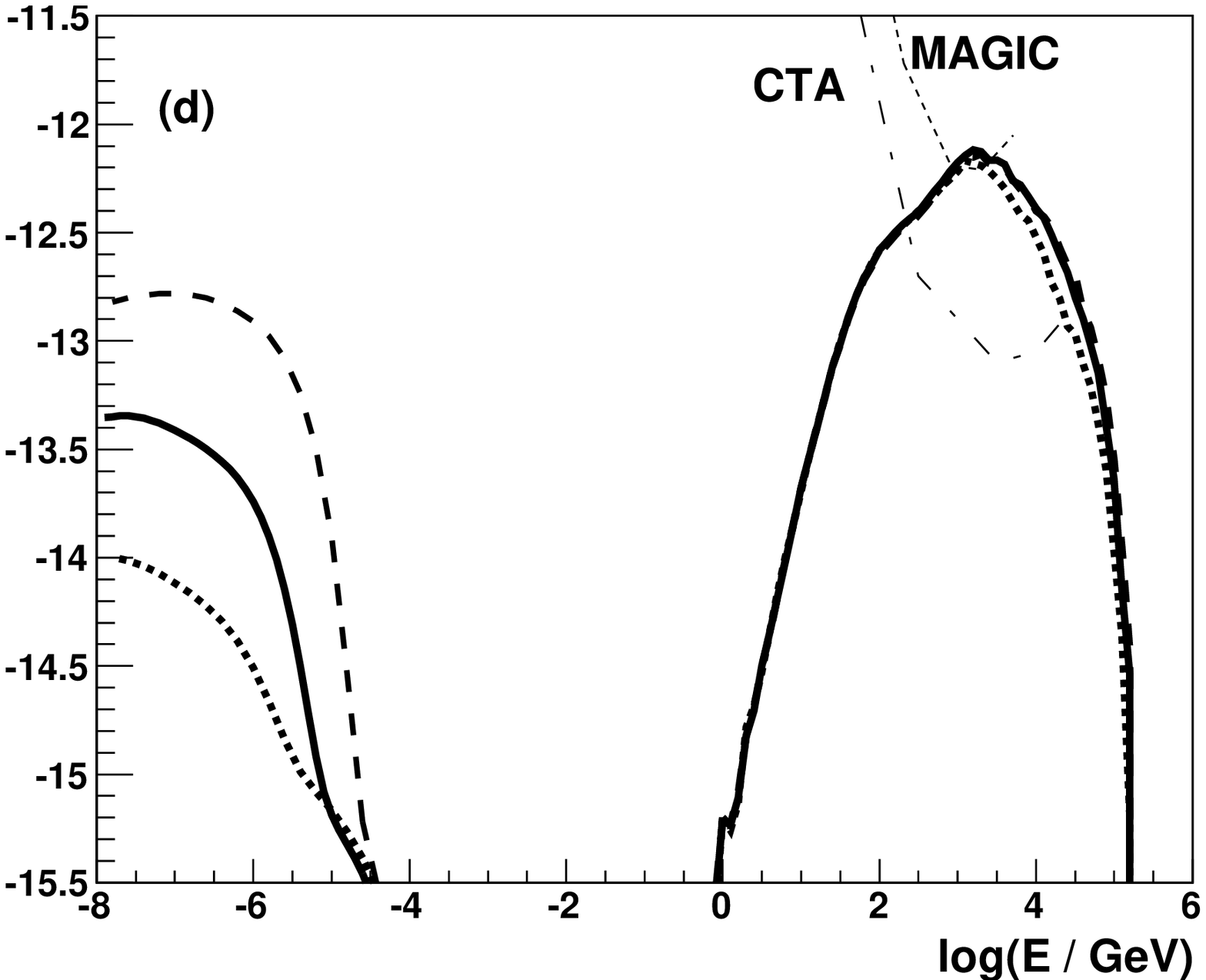}
\caption{Gamma-ray (IC) and X-ray (synchrotron) spectra (Spectral Energy Distribution - SED) produced in the nebula around the Black Widow binary system containing the millisecond pulsar B1957+20 for different model parameters. The spectra are produced by relativistic electrons which scatter the MBR and the infrared photons from the galactic disk. The maximum energies of the electrons are given by Eq.~\ref{eq4} and the minimum energies are equal to $E_{\rm w} = 3$ TeV. (a) Dependence of SED on the magnetization parameter $\sigma = 0.1$ (dashed), 0.01 (dotted), and 0.001 (solid) for the pulsar wind shock radius  $R_{\rm sh} = 10^{16}$ cm, the power law spectrum of the electrons with spectral index $\alpha = 2.5$ and the minimum magnetic field strength $B_{\rm min} = 0.5\mu$G.
(b) Dependence of SED on the radius of the pulsar wind shock $R_{\rm sh} = 10^{15}$ cm (dotted),
$10^{16}$ cm (solid), and $10^{17}$ cm (dashed), for  $\sigma = 0.01$, $\alpha = 2.5$, and 
$B_{\rm min} = 0.5\mu$G. (c) Dependence of SED on the spectral index of the electrons $\alpha = 2.1$ (dashed), 2.5 (solid), and 3 (dotted) for $R_{\rm sh} = 10^{16}$ cm,  $\sigma = 0.01$ and 
$B_{\rm min} = 0.5\mu$G. (d) Dependence of SED on the minimum value of the magnetic field
$B_{\rm min} = 0.5\mu$G (dotted), 1$\mu$G (solid), and 2$\mu$G (dashed) for $R_{\rm sh} = 10^{16}$ cm,  $\sigma = 0.01$, and $\alpha = 2.5$. It is assumed that the relativistic electrons take $10\%$ of the rotational energy lost by the pulsar.
The 100 hrs differential sensitivity of the MAGIC stereo system (thin dotted, Aleksic et al.~2012) and the 100 hrs CTA sensitivity (Actis et al.~2011) are also marked.
}
\label{fig2}
\end{figure*}
\section{Production of high energy radiation} 

We calculate the $\gamma$-ray spectra produced by relativistic electrons in the IC scattering of the MBR and the infrared radiation from the galactic disk. These electrons also produce synchrotron emission which can extend up X-ray energy range.
It is commonly expected that electrons accelerated at the pulsar wind shock obtain the power law spectrum. We assume that this spectrum has a lower energy cut-off at energies corresponding to the Lorentz factor of the pulsar wind, i.e $\gamma_{\rm w}$ is equal to a few times $10^6$. In our calculations we fix this value on 3 TeV, in agreement with 
the modelling of the PWNe (Kennel \& Coroniti~1984) and
recent calculations of the spectra of the electrons leaving the inner magnetospheres of the millisecond pulsars in the frame of the pair starved polar cap model (e.g. Zajczyk et al.~2010).
The electrons take a significant part of the energy lost by the millisecond pulsar, which is of the order of  $\sim 10\%$. The spectrum of the electrons extends up to the maximum energy described in Sect.~3. These electrons are accelerated close to the pulsar wind shock and diffuse to the outer region creating a tail trailing behind the pulsar.
In this calculations we take the energy density of the infrared galactic disk emission equal to 
$1.5$ eV cm$^{-3}$. It is assumed that  the magnetic field is enhanced by a factor of 3 in the pulsar wind shock and at larger distances continue to drop according to Eq.~3 up to the minimum value $B_{\rm min}$. This minimum magnetic field strength can be even below the magnetic field strength in the interstellar space (of the order of $\sim$2-6$\mu$G), since the volume of the pulsar wind nebula is not penetrated by the interstellar medium.

We assume that electrons are injected at the distance of the shock from the pulsar, $R_{\rm sh}$. They slowly diffuse outward according to the Bohm diffusion model in a decreasing magnetic field. During the diffusion process, the electrons interact with the background radiation producing 
GeV-TeV $\gamma$-rays in the IC process. We apply the Monte Carlo method in order to determine the energy of the $\gamma$-ray photons and the distance from the pulsar at which they are produced. For this purpose we modify the numerical code developed 
for the interaction and diffusion of electrons (Bednarek \& Sitarek~2007). This code allows us not only to calculate the spectrum of $\gamma$-rays produced by electrons but also determine their production sites around the pulsar, i.e. allowing us to study the morphology of the $\gamma$-ray source. Since the electrons are immersed in a relatively strong magnetic field, especially close to the pulsar wind shock, we also include in the simulations their synchrotron energy losses during the diffusion process. We calculate the X-ray spectra produced by these electrons in the synchrotron process.
In order to obtain reasonable precision of the IC $\gamma$-ray spectra, we simulate the propagation of $1.5\times 10^4$ electrons per decade of the spectrum. The spectra are obtained within different regions around the pulsar defined by the radius $R_{\rm Neb}$.

We investigate the dependence of the X-ray and $\gamma$-ray spectra on different parameters which determine the acceleration of the electrons (i.e. the magnetization parameter of the pulsar wind $\sigma$, the spectral index of the electrons' spectrum $\alpha$', the radius of the pulsar wind shock $R_{\rm sh}$; and the minimum value of the magnetic field in the nebula $B_{\rm min}$). As shown in Fig.~2, the TeV $\gamma$-ray spectra produced by the electrons in the IC process only weakly depend on the range of the considered parameters. On the other hand, the synchrotron X-ray emission depends on these parameters much stronger (intensity, shape, energy range). 
The strong dependence of the synchrotron emission is due to the strong dependence of the magnetic field in the vicinity of the pulsar
on the assumed parameters of the model. On the other hand, relatively weak dependence of the IC emission is due to the homogeneity of the 
background radiation field (MBR and infrared galactic background) which is up-scattered by the relativistic electrons.
We conclude that the TeV $\gamma$-ray fluxes expected in this model depend rather weakly on the details of the electron spectrum (in the considered range of parameters). However, their intensity is obviously determined by the energy conversion efficiency from the pulsar to the relativistic particles. In contrast, the spectra of the synchrotron radiation in the X-ray range much stronger depend on the spectrum of the electrons and the propagation model. 

We also investigate the $\gamma$-ray production in different volume around the Black Widow binary system B1957+20. 
The IC $\gamma$-ray and the synchrotron X-ray spectra are calculated assuming that this emission is produced within the region with the radius equal to 1.5 pc, 2.5 pc, 5 pc, 10 pc, and 15 pc (see Fig.~3). These dimensions correspond roughly to the angular size of the $\gamma$-ray source on the sky equal to 2, 3.4, 7, 14, and 20 arc min for the distance of the source equal to 2.5 kpc. The electrons expand into such a region due to their diffusion in the nebula. Moreover,
the TeV $\gamma$-ray source is also expected to be shifted from the present location of the Black Widow binary due to its motion and/or limitted in specific directions by the diffusion of the electrons confined by the presence of the bow shock. In the case of a source with the radius above $\sim5$ pc, the TeV $\gamma$-ray source should appear extended for the telescope array such as MAGIC. Our calculations show that most of the TeV $\gamma$-ray emission (i.e. within a factor of two) is already produced within a region with the radius of 5 pc. The shapes of the spectra, produced in specific parts of the $\gamma$-ray source, are quite similar since the background radiation field (MBR and infrared), scattered by the relativistic electrons, fills this region homogeneously. Moreover the cooling process of the electrons is not very efficient. The electrons do not usually interact frequently but in a specific interaction lose significant amount of their energy when producing TeV $\gamma$-rays. Due to the inefficient cooling, the parts of the spectra at low energies (in the GeV range), produced in the Thomson regime, are very similar.
On the other hand, the synchrotron X-ray emission does not depend on the considered radius of the source at energies above a few keV. This can be understood since the hard synchrotron radiation is mainly produced close to the pulsar wind shock within the region with the extend of $\sim$ 2 pc.
There is however an important contribution from the outer nebula to the part of the synchrotron spectrum at lower energies (below a few keV) since these electrons can still produce  keV photons in the assumed minimum magnetic field.

\begin{figure}
\vskip 6.6truecm
\includegraphics{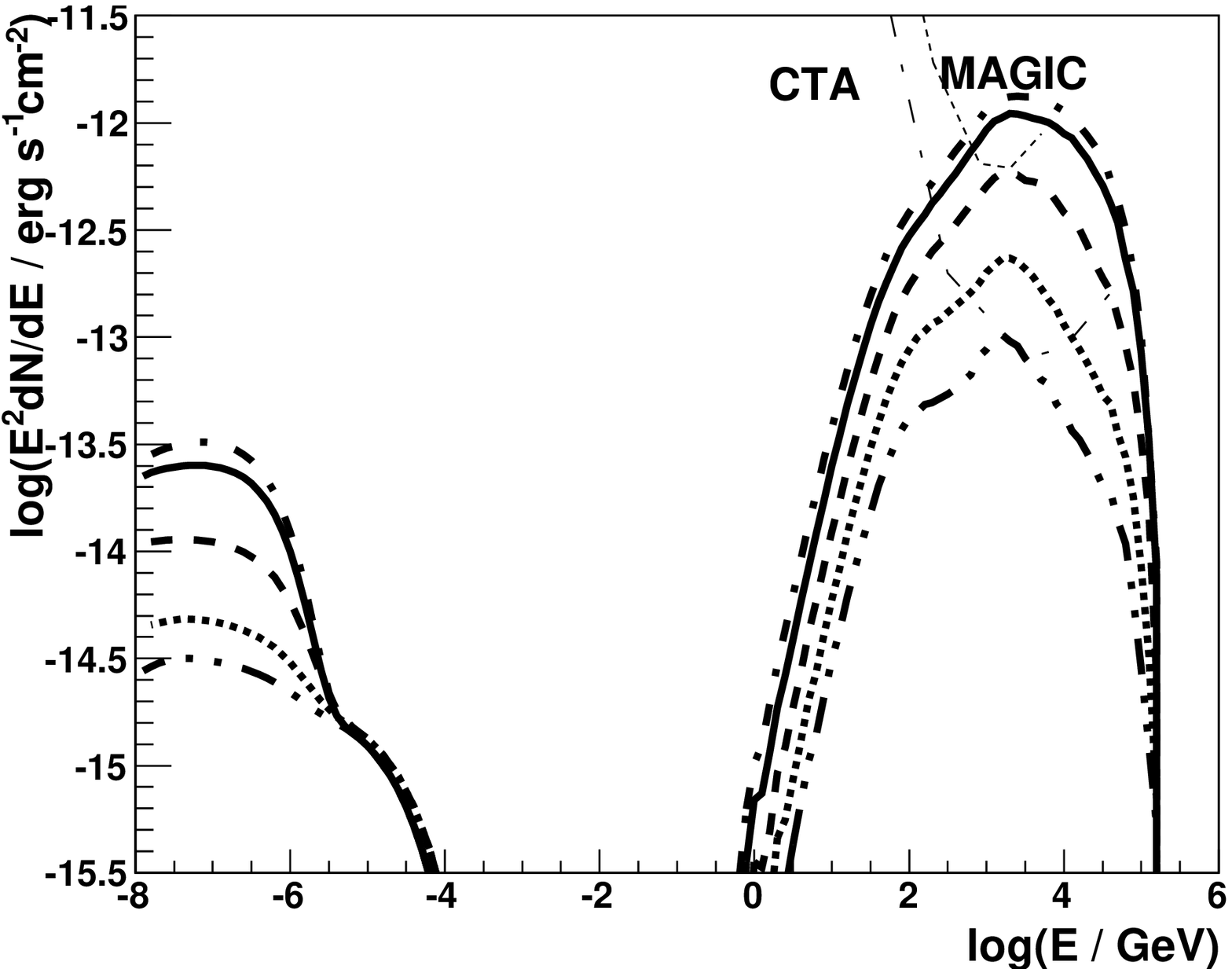}
\caption{SED of the $\gamma$-ray spectrum from the IC process and the synchrotron X-ray emission produced by relativistic electrons in the nebula around the Black Widow binary system B1957+20, integrated over a region with the radius: $R_{\rm Neb} = 1.5$ (dot-dot-dashed curve) pc, 2.5 pc (dotted), 5 pc (dashed), 10 pc (solid), and 15 pc (dot-dashed). The other parameters of the model are: $\sigma = 0.01$, the pulsar wind shock radius  $R_{\rm sh} = 10^{16}$ cm, the power law spectrum of the electrons with the spectral index $\alpha = 2.1$ between $E_{\rm w} = 3$ TeV and $E_{\rm max}$ (given by Eq.~\ref{eq4}), and the minimum magnetic field strength $B_{\rm min} = 0.5\mu$G.
The $\gamma$-ray spectra are produced by relativistic electrons which scatter the MBR and the infrared photons from the galactic disk. 
}
\label{fig3}
\end{figure}
\section{Comparison with observations of B1957+20} 

Finally, we compare the example calculations performed in terms of this modelling with the available observations of the Black Widow binary system B1957+20. The X-ray emission, extending along the direction of the motion of the binary, has been detected by {\it Chandra} (Stappers et al.~2003, Huang et al.~2012). The X-ray synchrotron emission expected in our model has to be consistent with this observed spectral features. Recently, the pulsed GeV $\gamma$-ray emission has been also reported from B1957+20 (Guillemot et al.~2012). The IC $\gamma$-ray emission, produced in the nebula by relativistic electrons, has to be below this pulsed emission. There are not available any positive detections or the upper limits on the TeV $\gamma$-ray emission from this source.

We have chosen intermediate parameters of the nebula from the range considered in Sect.~4.
The IC and the synchrotron spectra are confronted with the available observations of the binary system containing B1957+20 in Fig.~4. We have got good consistency with the level and shape of the X-ray spectrum from the nebula. Note that the X-ray observations put strong constraints on the 
parameters of the considered model. The emission extending up to $\sim$10 keV requires the presence of electrons with energies at least $\sim 4.7\times 10^5B_{\mu G}^{-1/2}$ GeV (see Eq.~\ref{eq5}).  On the other hand, the observed X-ray flux constrains the number of the relativistic electrons (which we fix on 10$\%$ of the pulsar energy loss rate) and the synchrotron energy loss rate which depends on $\propto B^2E_{\rm e}^2$.
Therefore, it is not so easy to model the observed X-ray spectrum correctly since the change of the parameters have strong effect on the energies and intensity of the emitted synchrotron radiation (see calculations in Fig.~2). We conclude that observed X-ray extended emission put strong constraints on the parameters of the considered model. Having obtained consistency with the observed synchrotron spectrum, we calculate the IC $\gamma$-ray spectrum for the same parameters (see the caption of Fig.~4). These spectra are confronted then with the sensitivities of the Cherenkov telescopes. We show the level of the IC emission expected from the region with the radius of $\sim$5 pc, which is shifted from the present location of the binary system by about the same distance in the direction opposite to the movement of the binary due to the motion of the binary system. Therefore, we conclude that the TeV $\gamma$-ray source should be extended for the Cherenkov telescopes.
The IC $\gamma$-ray spectrum is clearly above the 100 hr sensitivity of the future Cherenkov telescope Array (CTA). It is also on the 100 hr sensitivity limit of the MAGIC Cherenkov telescopes. 
We conclude that even with the present Cherenkov telescopes (MAGIC, VERITAS) the bow shock nebula around the Black Widow millisecond pulsar B1957+20 might be detected. 
Note however that as the source is expected to be extended, the sensitivity of the present Cherenkov telescopes might become worse than the point source sensitivity shown in Figs.~2-4.

\begin{figure}
\vskip 6.6truecm
\includegraphics{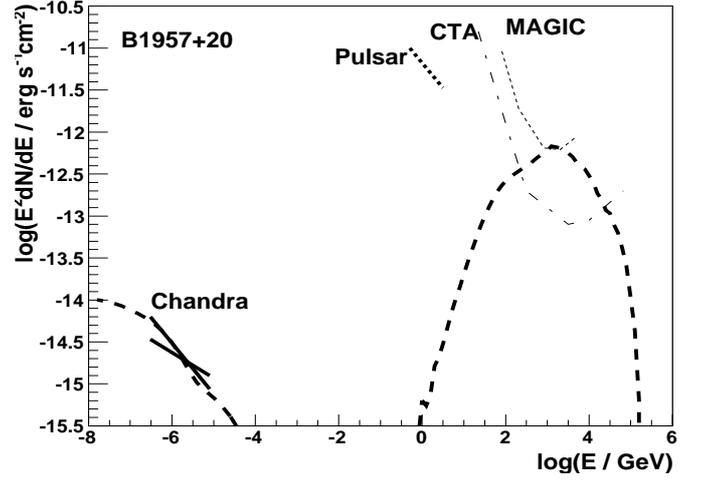}
\caption{The observations of the Black Widow binary containing millisecond pulsar B1957+20: the X-ray tail emission detected by {\it Chandra} (Huang et al.~2012) and the pulsed, phase averaged $\gamma$-ray emission discovered by {\it Fermi} (Guillemot et al.~2012) are compared with the calculations of the IC and the synchrotron emission by relativistic electrons in the nebula around this system. 
The $\gamma$-ray spectra are produced by the electrons which scatter both the MBR and the infrared photons from the galactic disk. The electrons have a power law spectrum with the index of $-2.1$ between $E_{\rm w} = 3$ TeV and $E_{\rm max} = 160$ TeV, given by Eq.~\ref{eq4} (thick dashed curves). The other parameters of the model are: the magnetization parameter $\sigma = 0.01$, the location of the shock $R_{\rm sh} = 10^{16}$ cm,
the minimum magnetic field strength $B_{\rm min} = 0.5\mu$G,  the confinement region of the electrons has the radius of $R_{\rm Neb} = 5$ pc. The 100 hrs differential sensitivity of the MAGIC stereo system is marked by the thin dotted curve (Aleksic et al.~2012)
and 100 hrs CTA sensitivity is marked by the thin dot-dashed curve (Actis et al.~2011).
}
\label{fig4}
\end{figure}
\section{Conclusion}

Assuming that the millisecond pulsars are able to accelerate electrons to relativistic energies in their vicinity, similarly 
as observed in the case of nebulae around classical pulsars, we calculate the synchrotron and the IC high energy emission from their nebulae. In fact, the existence of an extended synchrotron nebula
has been recently confirmed in the {\it Chandra} observations in the case of the Black Widow binary system containing millisecond pulsar B1957+20. Therefore, as an example, we consider the bow shock nebula around this object. Note that in contrast to the nebulae around classical pulsars, the soft radiation field in the nebula around B1957+20 is not dominated by the synchrotron radiation but by the MBR and infrared radiation from the galactic disk.
We have investigated the features of the X-ray and $\gamma$-ray spectra for likely range of parameters which determine the nebula, assuming that the propagation of electrons is determined by the diffusion process and/or the dynamical movement of the binary system. We conclude that the observed extended X-ray emission from the bow shock nebula can be explained by the synchrotron radiation of electrons provided that the energy conversion efficiency from the pulsar to the relativistic electrons is of the order of $10\%$. The TeV $\gamma$-ray emission, produced by the same electrons in the IC scattering process, is expected to be detectable by the future CTA instrument. The predicted emission is also on the level of the 100 hr sensitivity limit of the MAGIC telescopes. However, since the nebula is expected to be extended, due to rather slow cooling process of electrons, the detectability of the TeV $\gamma$-ray emission from the nebula around B1957+20 may be difficult. Note also that due to the motion of the binary system the 
TeV $\gamma$-ray nebula should be shifted in respect to the direction towards the binary system by the distance comparable to the extend of the source (see also Cheng et al.~2006).

Other bow shock nebulae around energetic pulsars should also emit synchrotron and IC high energy emission from their surrounding. However their detectability will strongly depend on the distance to the nebula. It can not be too large since the expected flux will be below detectability of the Cherenkov telescopes. But it should not be 
too close since the TeV $\gamma$-ray nebula will have very large dimensions on the sky which again will make problematic its detectability with the Cherenkov telescopes. For example, the bow shock nebula around nearby Geminga pulsar (at the distance 169 pc) may not be detected by the present Cherenkov telescopes. Due to its small distance, the angular size of the TeV nebula expected in terms of discussed above model, should be of the order of a few degrees, i.e  more in accordance with the recent report on the presence of the extended $\sim$20 TeV $\gamma$-ray source with diameter $(2.8\pm 0.8)^0$, towards the Geminga pulsar by the MILAGRO observatory (Abdo et al.~2009). However, such nebulae might be detected by the planned CTA which field of view can be as large as 8-9 degrees (Actis et al.~2011).

\begin{acknowledgements}
We would like to thank the Editor Steven N. Shore  and the Referee for useful comments.
This work is supported by the grants from the Polish MNiSzW through the NCN No. 2011/01/B/ST9/00411 and UMO-2011/01/M/ST9/01891. 
\end{acknowledgements}


\begin{thebibliography}{}

\bibitem{ab09} Abdo, A.A. et al. 2009 ApJ 700, L127 
\bibitem{ab10} Abdo, A.A. et al. 2010 ApJS 187, 460 
\bibitem{ac11} Actis, M. et al.. 2011 Exp.Astron. 32, 193
\bibitem{al12} Aleksic, J. et al. 2012 APh 35, 435
\bibitem{at93} Arons, J., Tavani, M. 1993 ApJ 403, 249
\bibitem{ar94} Arzoumanian, Z. et al. 1994 ApJ 426, L85
\bibitem{bs07} Bednarek, W., Sitarek, J. 2007 MNRAS 377, 920
\bibitem{br90} Brink, C. et al. 1990 ApJ 364, L37
\bibitem{bu96} Buccheri, R. et al. 1996 A\&AS 115, 305
\bibitem{ch06} Cheng, K.S., Taam, R.E., Wang, W. 2006 ApJ 641, 427
\bibitem{dh92} de Jager, O.C., Harding, A.K. 1992 ApJ 396, 161
\bibitem{fr88} Fruchter, A.S. et al. 1988 Nature 333, 237
\bibitem{fr95} Fruchter, A.S. et al. 1996 ApJ 443, 21
\bibitem{gu12} Guillemot, L. et al. 2012 ApJ 744, 33
\bibitem{hb07} Huang, H.H., Becker, W. 2007 A\&A 463, L5
\bibitem{hu12} Huang, R.H.H. et al. 2012 ApJ, in press (arXiv:1209.5871)
\bibitem{hu11} Hui, C.Y. et al. 2011 ApJ 726, 100
\bibitem{kc84} Kennel, C.F., Coroniti, F.V. 1984 ApJ 283, 694
\bibitem{kh88} Kulkarni, S.R., Hester, J.J. 1988 Nature, 335, 801
\bibitem{re07} Reynolds, M.T. et al. 2007 MNRAS 379, 1117
\bibitem{sd03} Sefako, R.R., de Jager, O.C. 2003 ApJ 593, 1013
\bibitem{st03} Stappers, B.W. et al. 2003 Science 299, 1372
\bibitem{sm98} Strong, A.W., Moskalenko, I.V. 1998 ApJ 509, 212
\bibitem{ta12} Takata, J., Cheng, K.S., Taam, R.E. 2012 ApJ 745, 100
\bibitem{vk11} van Kerkwijk, M.H. et al. 2011 ApJ 728, 95
\bibitem{vp88} van paradijs, J. et al. 1988 Nature 334, 684
\bibitem{wu12} Wu, E.M.H. et al. 2012 ApJ, in press (arXiv:1210.7209)
\bibitem{za10} Zajczyk, A. et al. 2010, in Proc. High Time Resolution Astrophysics - The Era of Extremely Large Telescopes Agios Nikolaos (Crete Greece), Procceedings of Science published on line: http://pos.sissa.it/cgi-bin/reader/conf.cgi?confid=108, id.52

\end{thebibliography}
\end{document}